

\font\mbf=cmbx10 scaled\magstep1

\def\bs{\bigskip}
\def\ms{\medskip}

\def\ni{\noindent}
\def\cl{\centerline}

\def\title#1{\cl{\mbf #1}}
\def\ref#1#2#3#4{#1\ {\it#2\ }{\bf#3\ }#4\par}
\def\refb#1#2#3{#1\ {\it#2\ }#3\par}
\def\CQG{Class.\ Qu.\ Grav.}
\def\GRG{Gen.\ Rel.\ Grav.}
\def\IJMP{Int.\ J.\ Mod.\ Phys.}
\def\PR{Phys.\ Rev.}
\def\PRS{Proc.\ Roy.\ Soc.\ Lond.}

\def\d{\hbox{d}}
\def\f#1#2{{\textstyle{#1\over#2}}}

\magnification=\magstep1
\overfullrule=0pt

\title{Signature change at material layers and step potentials}
\bs\cl{\bf Sean A. Hayward}
\ms\cl{Department of Physics, Kyoto University, Kyoto 606-01, Japan}
\ms\cl{\tt hayward@murasaki.scphys.kyoto-u.ac.jp}
\bs\ni
{\bf Abstract.}
For a contravariant 4-metric which changes signature
from Lorentzian to Riemannian across a spatial hypersurface,
the mixed Einstein tensor is manifestly non-singular.
In Gaussian normal coordinates, the metric contains a step function
and the Einstein tensor contains the Dirac delta function
with support at the junction.
The coefficient of the Dirac function is a linear combination of
the second fundamental form (extrinsic curvature) of the junction.
Thus, unless the junction has vanishing extrinsic curvature,
the physical interpretation of the metric is
that it describes a layer of matter (with stresses but no energy or momentum)
at the junction.
In particular, such metrics do not satisfy the vacuum Einstein equations,
nor the Einstein-Klein-Gordon equations and so on.
Similarly, the d'Alembertian of a Klein-Gordon field contains
the Dirac function with coefficient given by the momentum of the field.
Thus, if the momentum of the field does not vanish at the junction,
the physical interpretation is
that there is a source (with step potential) at the junction.
In particular, such fields do not satisfy the massless Klein-Gordon equation.
These facts contradict claims in the literature.
\bs\cl{PACS: 04.20.Cv, 02.40.Ky}
\bs\ni
Recently there has been growing interest in metrics which change signature,
particularly from Riemannian to Lorentzian across a spatial hypersurface.
Such metrics can be thought of as
classical idealisations of tunnelling solutions in quantum cosmology,
where the idea originally arose [1].
In order to satisfy the Einstein equations,
the junction of signature change must be totally geodesic,
i.e.\ have vanishing second fundamental form (extrinsic curvature) [1].
More generally, the momenta of the physical fields must vanish at the junction.
Although this is well known in some quarters,
much of the recent literature on signature change
contains claims to the contrary.

In particular,
Ellis et al.\ and subsequent authors [2--11] have claimed that
there are signature-changing solutions to the Einstein equations,
without surface layers,
with non-vanishing extrinsic curvature at the junction.
The incorrectness of such claims will be demonstrated here
simply by calculating the Einstein tensor of the proposed solutions,
in the coordinates that Ellis et al.\ used.

Also, Dray et al.\ and subsequent authors [12--18] have claimed that
there are solutions to the massless Klein-Gordon equation
across a change of signature with non-vanishing momentum at the junction.
The incorrectness of such claims will be demonstrated here
simply by calculating the d'Alembertian of the proposed solutions,
in the same coordinates as above.

These points have mostly been made before [19--30]
but it is worth repeating them
since many authors continue to claim the contrary.
Moreover, rather than just showing that
the proposed solutions do not satisfy the equations,
this article will go a stage further
and exhibit field equations which they do satisfy,
namely the Einstein or Klein-Gordon equations with certain sources.
This explains the physical meaning of the proposed solutions:
they describe a source layer at the junction of signature change.
In the Einstein case there is a layer of matter at the junction,
and in the Klein-Gordon case there is a step potential across the junction.
This is most easily seen in a new representation presented here.

Consider first the simplest example,
the massless Klein-Gordon equation for a 1-dimensional contravariant metric
$$\gamma=-\lambda{\partial\over{\partial t}}\otimes{\partial\over{\partial t}}.
\eqno(1)$$
The d'Alembertian $\Delta$ of this metric is given by
$$\Delta\phi=-\lambda\phi''-\f12\lambda'\phi'\eqno(2)$$
where the prime denotes differentiation with respect to $t$.
Consider the case
$$\lambda=\varepsilon\eqno(3)$$
where $\varepsilon$ is the sign of $t$.
This describes a metric which changes signature at $t=0$,
such that $t$ is proper time or distance either side of the junction $t=0$.
Note that
$$\varepsilon'=2\delta\eqno(4)$$
where $\delta$ is the Dirac delta function in $t$ with support at $t=0$.
Both $\varepsilon$ and $\delta$ are distributions
in the sense of Schwartz [31--32].
Therefore the d'Alembertian is given by
$$\Delta\phi=-\varepsilon\phi''-\delta\phi'\eqno(5)$$
which is evidently a Schwartz distribution
if $\phi$ is $C^1$ at the junction and $C^2$ elsewhere.
Thus solutions of the massless Klein-Gordon equation
must have vanishing momentum $\phi'$ at the junction:
$$\Delta\phi=0\qquad\Rightarrow\qquad\phi'|_{t=0}=0.\eqno(6)$$
This is the junction condition
for the Klein-Gordon field at a change of signature.
It follows that the general solution to the equation
is just $\phi=\phi_0$ for constant $\phi_0$.

Several authors have claimed otherwise [12--18].
The most common claim is that
$$\phi=At+\phi_0\eqno(7)$$
satisfies the massless Klein-Gordon equation,
sometimes expressed in the same coordinates, sometimes not.\footnote\dag
{Dray et al.\ [12--14] also suggest $\phi=A|t|+\phi_0$ as solutions
(in different coordinates)
and make contradictory statements about which set they prefer [25].}
Evidently this claim is incorrect:
the d'Alembertian of (7) is
$$\Delta(At+\phi_0)=-A\delta.\eqno(8)$$
The problem was that none of the authors actually calculated the d'Alembertian
in a well defined form,
but instead solved the equation either side of the junction
and joined the solutions together according to a continuity condition.
Clearly such a procedure will miss the $\delta$ term in $\Delta$.

As to the physical meaning of the proposed solutions (7),
they do satisfy the Klein-Gordon equation with potential,
$$\Delta\phi={\partial V\over{\partial\phi}}\eqno(9)$$
for the step potential
$$V=-\f12A^2\hbox{sign}\left({\phi-\phi_0\over{A}}\right).\eqno(10)$$
There seems to be no good reason to accept such a step potential
at a change of signature,
particularly one so finely tuned to the proposed solution.

Actually, the proposed solutions (7) formally satisfy
the Klein-Gordon equation (9) with step potential (10) in any coordinates,
including the above $\lambda=\varepsilon$ coordinates
and the $\lambda=t^{-1}$ coordinates that Dray et al.\ [12--14] prefer.
This happens because (9) may be formally integrated as an exact differential
for any choice of $\lambda$:
$$V=\int\Delta\phi\,\d\phi=-\int(\lambda\phi''+\f12\lambda'\phi')\phi'\d t
=-\f12\int(\lambda(\phi')^2)'\d t=-\f12\lambda(\phi')^2
=\f12\gamma(\d\phi,\d\phi).\eqno(11)$$
The last expression is manifestly coordinate-independent.
The point is that, in whatever coordinates,
these proposed solutions do not satisfy the massless Klein-Gordon equation,
but instead correspond to a step potential.

The above point is important because
some of the authors who proposed the invalid solutions are still in denial,
claiming that the proposed solutions do satisfy
some form of the massless Klein-Gordon equation in some sense,
to be determined by sufficient imagination [7--8,12--14].
The futility of these efforts should now be clear:
the proposed solutions do satisfy the Klein-Gordon equation with step
potential,
therefore they do not satisfy the Klein-Gordon equation with constant
potential.

Consider now the second example,
the Einstein equation for a 4-manifold
with contravariant metric $\gamma^{\mu\nu}$.
Take a coordinate system $x^\mu=(x^0,x^i)$
with $x^0=t$ labelling a family of hypersurfaces
and $x^i$ being coordinates on the hypersurfaces.
Denote the components of the metric by
$$\eqalignno
{&\gamma^{00}=-\lambda&(12a)\cr
&\gamma^{0i}=\lambda s^i&(12b)\cr
&\gamma^{ij}=h^{ij}.&(12c)\cr}$$
In the usual terminology, $h_{ij}$ is the induced metric on the hypersurfaces,
assumed spatial (positive definite),
$s^i$ is the shift vector and $\lambda$ is the inverse squared lapse function;
that is, if $\lambda>0$ then the signature of the metric is Lorentzian
and the usual lapse function is $\lambda^{-1/2}$.
If $\lambda<0$ then the signature is Riemannian (Euclidean).
A change of signature may be described by
$\lambda$ changing sign across one of the hypersurfaces, say $t=0$.
Whatever the signature,
the mixed Einstein tensor $G_\mu^\nu$ of $\gamma^{ij}$ may be calculated to be
$$\eqalignno
{&G_0^0=-\lambda G_{00}+\lambda s^iG_{0i}&(13a)\cr
&G_0^i=h^{ij}G_{0j}+\lambda s^iG_{00}&(13b)\cr
&G_i^0=-\lambda G_{0i}+\lambda s^jG_{ij}&(13c)\cr
&G_i^j=h^{jk}G_{ik}+\lambda s^jG_{0i}&(13d)\cr}$$
where
$$\eqalignno
{&\lambda G_{00}=\f12 R
-\f18\lambda(h^{ik}h^{jl}-h^{ij}h^{kl})(L_{v-s}h_{ij})L_{v-s}h_{kl}&(14a)\cr
&G_{0i}=\f12(h_i^kh^{jl}-h_i^jh^{kl})D_jL_{v-s}h_{kl}
+\f14(h_i^kh^{jl}-h_i^jh^{kl})(L_{v-s}h_{kl})\lambda^{-1}D_j\lambda&(14b)\cr
&G_{ij}=E_{ij}+\f12\lambda(h_i^kh_j^l-h_{ij}h^{kl})L_{v-s}L_{v-s}h_{kl}\cr
&\qquad\qquad+\f12\lambda(-h_i^kh_j^mh^{ln}+\f12h_i^kh_j^lh^{mn}
+\f34h_{ij}h^{km}h^{ln}-\f14h_{ij}h^{kl}h^{mn})(L_{v-s}h_{kl})L_{v-s}h_{mn}\cr
&\qquad\qquad+\f12(h_i^kh_j^l-h_{ij}h^{kl})\lambda^{-1}D_kD_l\lambda
-\f34(h_i^kh_j^l-h_{ij}h^{kl})\lambda^{-2}(D_k\lambda)D_l\lambda\cr
&\qquad\qquad+\f14(L_{v-s}\lambda)(h_i^kh_j^l-h_{ij}h^{kl})L_{v-s}h_{kl}
&(14c)\cr}$$
where $E_{ij}$, $R$ and $D_i$ are respectively
the Einstein tensor, Ricci scalar and covariant derivative of $h_{ij}$,
$L_u$ denotes the Lie derivative along a vector $u$,
and $v=\partial/\partial t$.
In a Riemannian region, one may take Gaussian normal coordinates:
$(\lambda,s)=(-1,0)$.
Similarly, in a Lorentzian region, one may take normal coordinates:
$(\lambda,s)=(1,0)$.
So a change of signature at $t=0$
may be described by taking coordinates such that
$$\eqalignno{&\lambda=\varepsilon&(15a)\cr &s^i=0&(15b)\cr}$$
where $\varepsilon$ is the sign of $t$.
This means that $t$ is the proper time or distance
normal to the junction of signature change.
These are the coordinates favoured by Ellis et al.\ [2].
It is crucial to note that
$$L_v\varepsilon=2\delta\eqno(16)$$
where $\delta$ is the Dirac delta function in $t$ with support at $t=0$.
Both $\varepsilon$ and $\delta$ are distributions in the sense of Schwartz.
Direct substitution of this coordinate choice into (13--14)
yields the mixed Einstein tensor as
$$\eqalignno
{&G_0^0=-\f12 R
+\f18\varepsilon(h^{ik}h^{jl}-h^{ij}h^{kl})(L_vh_{ij})L_vh_{kl}&(17a)\cr
&G_0^i=\f12(h^{ik}h^{jl}-h^{ij}h^{kl})D_jL_vh_{kl}&(17b)\cr
&G_i^0=-\f12\varepsilon(h_i^kh^{jl}-h_i^jh^{kl})D_jL_vh_{kl}&(17c)\cr
&G_i^j=E_i^j+\f12\varepsilon(h_i^kh^{jl}-h_i^jh^{kl})L_vL_vh_{kl}\cr
&\qquad\qquad+\f12\varepsilon(-h_i^kh^{jm}h^{ln}+\f12h_i^kh^{jl}h^{mn}
+\f34h_i^jh^{km}h^{ln}-\f14h_i^jh^{kl}h^{mn})(L_vh_{kl})L_vh_{mn}\cr
&\qquad\qquad+\f12\delta(h_i^kh^{jl}-h_i^jh^{kl})L_vh_{kl}&(17d)\cr}$$
which is evidently a Schwartz distribution [31--32]
if $h_{ij}$ is $C^1$ in $t$ at the junction and $C^2$ otherwise.
It contains the Dirac function $\delta$,
with the coefficient being a linear combination of
the second fundamental form (extrinsic curvature) $L_vh_{ij}$.
Therefore the vacuum Einstein equations
require the extrinsic curvature of the junction to vanish:
$$G_\mu^\nu=0\qquad\Rightarrow\qquad L_vh_{ij}|_{t=0}=0.\eqno(18)$$
The same condition holds for the Einstein-Klein-Gordon equations,
the Einstein-Maxwell equations, the Einstein-fluid equations and so on.
These are the junction conditions for the Einstein field
at a change of signature.
The Einstein equations divide into three parts:
the usual Einstein equations in the Lorentzian region,
the corresponding Riemannian Einstein equations in the Riemannian region,
and the above junction conditions at the junction.

When Ellis et al.\ [2] performed the above calculation
in the homogeneous isotropic case,
they missed the term in $\delta$.
Subsequently they solved the resulting equations---the Einstein equations
away from the junction---for various stress-energy tensors $T_{\mu\nu}$,
producing what they called solutions to the Einstein equations
which do not satisfy the junction conditions.
Clearly this is incorrect;  for such a proposed solution,
$$G_\mu^\nu-T_\mu^\nu
=\f12\delta(h_\mu^ih^{j\nu}-h_\mu^\nu h^{ij})L_vh_{ij}\eqno(19)$$
whereas it should be zero.
Moreover, the physical meaning of such metrics is now clear:
they describe a layer of matter at the junction.
That is, the right-hand side of (19) represents
a stress-energy tensor with support at the junction.
This is the physical interpretation of the Einstein equation.

This matter has various unphysical properties.
Since it has support on a spatial hypersurface, it is non-causal or tachyonic.
It represents the instantaneous creation and destruction of matter.
It violates conservation of matter.
It violates all the usual energy conditions
because in the normal coordinates,
it has vanishing energy and momentum but non-vanishing stresses.
In short, there is no good reason to accept such a material source.

It should be noted that all previous work on signature change [1--30]
concerns a covariant metric rather than a contravariant metric.
Specifically, using the inverse squared lapse function $\lambda$
effects a contravariant representation,
whereas using the squared lapse function $\nu=\lambda^{-1}$
(or the complex lapse function $\sqrt\nu$)
effects a covariant representation.
When using $\nu$, one may take either discontinuous $\nu=\varepsilon$
or smooth $\nu$, with the canonical choice $\nu=t$ [26--29].
The discontinuous and smooth covariant representations
yield the same class of solutions,
related by coordinate transformations.\footnote*
{The claim to the contrary by Kossowski \& Kriele [27] is incorrect.
The problem was that, in their definition of the discontinuous representation,
they demanded that $h_{ij}$ be $C^2$.
This is inconsistent with generic solutions, since by (17d),
if $E_i^j-G_i^j$ is continuous and non-zero at the junction,
then $L_vL_vh_{ij}$ is discontinuous.}
Similarly, when using $\lambda$,
one may take either discontinuous $\lambda=\varepsilon$, as in this article,
or smooth $\lambda$, with the canonical choice $\lambda=t$.
The latter representation will be discussed elsewhere,
as will the relationship between the four representations:
covariant and contravariant, smooth and discontinuous.
The class of solutions is the same,
related by the coordinate transformations
$$t_\lambda=\f14\varepsilon t_\varepsilon^2=\f19t_\nu^3\eqno(20)$$
where $t_\lambda$ and $t_\nu$ are the canonical times
in the contravariant and covariant smooth representations respectively,
and $t_\varepsilon$ is the time
in either discontinuous representation.\footnote\ddag
{Note that the time $t$ is real in all four representations,
and in all the approaches examined in my papers [19--25].
I mention this since some authors [8]
insist that my approaches involve imaginary time
and are therefore irrelevant to theirs.}

The point is that these are just different representations of the same theory,
with equivalent solutions.
One may use whatever representation is convenient.
The discontinuous contravariant representation has been used here
because it has the advantage that
the physical meaning of the invalid proposed solutions is clearest.

This happens because of technical problems with the other representations.
Recall that the mixed Einstein tensor (17)
contains the distributions $\varepsilon$ and $\delta$ linearly
in the discontinuous contravariant representation.
In the discontinuous covariant representation,
these expressions become $1/\varepsilon$ and $-\delta/\varepsilon^2$,
which require interpretation.
Similarly, in both of the smooth representations,
substituting the proposed invalid solutions into the Einstein tensor
yields expressions involving not $\delta$ but negative powers of $t$,
again requiring interpretation.
Some authors [7--14] apparently believe that such expressions cancel to zero,
or being ambiguous, may be set to zero.
A similar technical problem provides the reason for taking
the mixed rather than covariant or contravariant Einstein tensor:
in the discontinuous contravariant representation,
$G^0_0$ is a Schwartz distribution, but $G_{00}$ and $G^{00}$
contain respectively $\varepsilon^{-1}$ and $\varepsilon^2$,
again requiring interpretation.
These technical problems with other representations
are all rather minor and easily resolved.
Yet they appear to have allowed confusion.
The point is that such problems do not occur at all
in the representation adopted in this article.

The above remarks are relevant because
there is considerable disinformation in the literature [2--18]
to the effect that the Einstein (or Klein-Gordon) equation does admit
signature-changing solutions with non-vanishing momentum at the junction,
if one uses certain coordinates or certain forms of the equation,
or solves the equation in a certain way,
or adopts a certain philosophy.\footnote\S
{Recently a more radical philosophy has also arisen,
namely that it is better not to satisfy the field equations [6,9].
The implication is that this was the intention all along.}
These arguments are all fallacious [19--25]
but are being pursued so widely and repeatedly
as to confuse the general audience.
Nevertheless,
such arguments are irrelevant to the main points of this article, as follows.

If a signature-changing solution
to the Einstein (or Klein-Gordon) equation is proposed,
it may be transformed to the discontinuous contravariant representation,
if it is not already so expressed.
The mixed Einstein tensor (or d'Alembertian) of the proposed solution
may then be calculated.
It is manifestly well defined
for the type of proposed solutions discussed above,
containing the Dirac delta function with support at the junction,
unless the extrinsic curvature (or Klein-Gordon momentum)
vanishes at the junction.
Thus proposed solutions not satisfying these junction conditions
actually do not satisfy
the vacuum Einstein (or massless Klein-Gordon) equation.
Instead, they satisfy the Einstein equation
with a certain delta-function stress-energy tensor
(or the Klein-Gordon equation with a certain step potential).
Thus the physical meaning of such proposed solutions is now clear:
they describe a layer of matter (or a step potential) at the junction.
\bs\ni
Research supported by the Japan Society for the Promotion of Science.
\bs
\begingroup
\parindent=0pt\everypar={\global\hangindent=20pt\hangafter=1}\par
{\bf References}\par
\ref{[1] Gibbons G W \& Hartle J B 1990}\PR{D42}{2458}
\ref{[2] Ellis G, Sumeruk A, Coule D \& Hellaby C 1992}\CQG{9}{1535}
\ref{[3] Ellis G F R 1992}\GRG{24}{1047}
\ref{[4] Kerner R \& Martin J 1993}\CQG{10}{2111}
\ref{[5] Ellis G F R \& Piotrkowska K 1994}\IJMP{D3}{49}
\ref{[6] Carfora M \& Ellis G 1995}\IJMP{D4}{175}
\ref{[7] Hellaby C \& Dray T 1994}\PR{D49}{5096}
\ref{[8] Hellaby C \& Dray T 1995}
{Comparison of approaches to classical signature change, \PR}D{(to appear)}
\ref{[9] Embacher F 1995}\PR{D51}{6474}
\ref{[10] Embacher F 1995}\PR{D52}{2150}
\ref{[11] Embacher F 1995}\CQG{12}{1723}
\ref{[12] Dray T, Manogue C A \& Tucker R W 1991}\GRG{23}{967}
\ref{[13] Dray T, Manogue C A \& Tucker R W 1993}\PR{D48}{2587}
\refb{[14] Dray T, Manogue C A \& Tucker R W 1995}
{Boundary conditions for the scalar field in the presence of signature change}
{(gr-qc/9501034)}
\ref{[15] Romano J D 1993}\PR{D47}{4328}
\ref{[16] Alty L J 1994}\CQG{11}{2523}
\refb{[17] Alty L J \& Fewster C J 1995}
{Initial value problems and signature change}{(gr-qc/9501026)}
\refb{[18] Egusquiza I L 1995}
{Self-adjoint extensions and signature change}{(gr-qc/9503015)}
\ref{[19] Hayward S A 1992}\CQG9{1851; erratum 2453}
\refb{[20] Hayward S A 1992}
{Comment on ``Change of signature in classical relativity''}{}
\ref{[21] Hayward S A 1993}\CQG{10}{L7}
\refb{[22] Hayward S A 1993}
{Junction conditions for signature change}{(gr-qc/9303034)}
\ref{[23] Hayward S A 1994}\CQG{11}{L87}
\ref{[24] Hayward S A 1995}{Comment on ``Failure of standard conservation laws
at a classical change of signature'', \PR}D{(to appear)}
\refb{[25] Hayward S A 1995}{Comment on ``Boundary conditions for the scalar
field in the presence of signature change''}{(gr-qc/9502001)}
\ref{[26] Kossowski M \& Kriele M 1993}\CQG{10}{1157}
\ref{[27] Kossowski M \& Kriele M 1993}\CQG{10}{2363}
\ref{[28] Kossowski M \& Kriele M 1994}\PRS{A444}{297}
\ref{[29] Kossowski M \& Kriele M 1995}\PRS{A446}{115}
\ref{[30] Kriele M \& Martin J 1995}\CQG{12}{503}
\refb{[31] Schwartz L 1950}{Th\'eorie des Distributions}{(Paris: Hermann)}
\refb{[32] Gel'fand I M \& Shilov G E 1964}{Generalized Functions}
{(New York: Academic Press)}
\endgroup
\bye